\documentclass[aps,twocolumn,prd,nofootinbib]{revtex4}
\usepackage{amsmath}
\usepackage{graphicx}
\usepackage{dcolumn}
\usepackage{bm}
\usepackage{amssymb}
\usepackage{latexsym}

\def\be{\begin{equation}}
\def\ee{\end{equation}}
\def\bea{\begin{eqnarray}}
\def\eea{\end{eqnarray}}

\begin{document}

\title{Dark Energy and Its Interactions with Neutrinos}

\author{Xinmin Zhang}
\affiliation{Institute of High Energy Physics, Chinese
Academy of Sciences, P.O. Box 918-4, Beijing 100049, P. R. China}

\begin{abstract}
In this talk I will firstly review on the current constraints on
the equation of state of the dark energy from observational data,
then present a new scenario of dark energy dubbed {\it
Quintom}. The recent fits to the type Ia supernova data and
the cosmic microwave background and so on in the literature find that the behavior of dark energy is
to great extent in consistency with a cosmological constant,
however the dynamical dark energy scenarios are generally not
ruled out, and one class of models with an equation of state
transiting from below $-1$ to above $-1$ as the redshift increases is
mildly favored. The second part of the talk is on interacting dark
energy. I will review briefly on the models of neutrino dark
energy.

\end{abstract}

\maketitle

\section{Introduction}

The recent observational data from type Ia supernova (SNIa), cosmic
microwave background (CMB) radiation and large scale structure
(LSS) have provided strong evidences for a spatially flat and
accelerated expanding universe at the present time. In the context
of Friedmann-Robertson-Walker cosmology, this acceleration is
attributed to the domination of a component, dubbed dark energy.
The simplest candidate for dark energy seems to be a remnant small
cosmological constant. However, many physicists are attracted by
the idea that dark energy is due to a dynamical component, such as
a canonical scalar field $Q$\cite{Quintessence}, named {\it
Quintessence}. In this paper I will show firstly that the recent
fits to the SNIa, CMB data and so on in the literature find that the
behavior of dark energy is to great extent in consistency with a
cosmological constant, however the dynamical dark energy scenarios
are generally not ruled out and in fact one class of models with
an equation of state (EOS) transiting from below $-1$ to above
$-1$\cite{Quintom} as the redshift increases, {\it Quintom} is
mildly favored\cite{FWZ,Qx,guo}.

The second part of this talk is on interacting dark energy. Being
a dynamical component, the scalar field of dark energy is expected to
interact with the ordinary matters. There are many discussions on
the explicit couplings of Quintessence to baryons, dark matter and
photons, however for most of the cases the couplings are strongly
constrained. But still there are exceptions. In this paper I will
review about the recent studies on the models of neutrino dark
energy. The paper is organized as follows: in section II I will
review briefly about the current constraints on the dark energy
from observational data; in section III I will review on the
Quintom scenario of dark energy and in section IV I will study the
neutrino dark energy models. The section V is the discussion and
the summary of this talk.

\section{Current constraints on the equation of state of the dark energy}

The quantity which characterizes the properties of the dark energy
models is the equation of state defined as the ratio of the
pressure to the energy density: $w = P/\rho$. For
instances in models of dark energy provided by the (true or false)
vacuum energy the equation of state $w$ is a constant and equals
to $-1$; in theories of Quintessence $w$ varies as a function of
redshift,
 however no matter how it evolves $w$ is restricted to be
larger than $-1$; in theories of Phantom\cite{Phantom} where the
kinetic term has a negative sign $w$ is smaller than $-1$ and no
matter how it evolves it will never cross over the cosmological
constant boundary into the regime of Quintessence.

In a model independent way to constrain the dark energy, one
usually considers some kind of parameterizations of the equation
of state. For instance, one possibility is expanding the equation
of state in the powers of the redshift. For small redshift one has
\begin{equation}
w(z)=W_0+W_1 z.
\end{equation}
Another type of parametrization was proposed
by Ref. \cite{linder}:
\begin{equation}
w(z)=W_1+W_a z/(1+z)~.
\end{equation}

Using the "gold" set of 157 SNIa published by Riess et al. in
\cite{Riess04} many papers in the literature have presented the
analysis on the constraints on the equation of state of the dark
energy. The results in general show that the best fitting model of
dark energy is the one with an equation of state
 transiting from below $-1$ to above $-1$ as the
redshift increases, however at $2 \sigma$ the cosmological
constant fits well to the data\cite{FWZ, Quintom}.
\begin{figure}[htbp]
\includegraphics[scale=0.6]{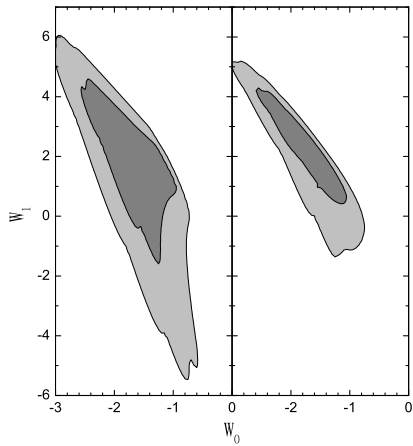}
\caption{SNIa and CMB constraints on the oscillating dark energy
model\cite{xia}.  The best fit values are shown in the centers of each
panel. The grey and light grey areas show the $1\sigma$ and
$2\sigma$ confidence regions respectively. \label{fig:osc}}
\end{figure}

\begin{figure}[htbp]
\begin{center}
\includegraphics[scale=0.5]{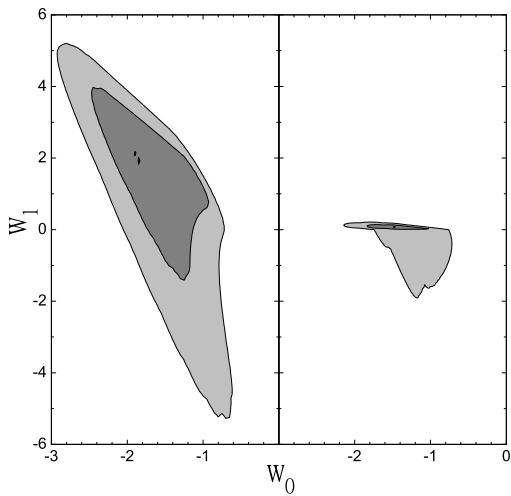}
\caption{SNIa and CMB  constraints on the linear parametrized
dark energy model\cite{xia}. Left : SNIa only; Right: SNIa $+$ CMB
constraints. \label{fig:lin}}
\end{center}
\end{figure}

Both parameterizations above make good approximations to probe the
behavior of dark energy around the present epoch, while the former
model leads to poor parameterization at very large redshift. As
one example, in Ref.\cite{xia} we have considered one type of
parametrization, the oscillating Quintom
\begin{equation}
w(z) = W_0 + W_1 \sin z.
\end{equation}
The oscillating Quintom in (3) differs from the model with a
linearly parametrized equation of state in (1) at high redshift,
however at low redshift coincides with it. Fig.1 shows the
constraints on the parameters of oscillating Quintom from SNIa and
CMB. For comparison in Fig.2 we show the corresponding constraints on
the parameters $( W_0, ~~ W_1)$ given in (1). One can see the
similarity when considering only the SNIa data and the significant
difference when including the CMB data. This exercise indicates
the dependence of the "global fitting" results on the ways of
parameterizing the equation of state of dark energy.

Another comment on the "global fitting" results is that the
previous fittings in the literature have simply fully or partially
neglected the perturbation of dark energy with an equation of
state acrossing $-1$. Recently with my colleagues I have developed a
consistent way to include the perturbation of dark energy with
equation of state crossing $-1$  and we have shown that in
general the parameter space will get enlarged when including the
perturbation than switching the dark energy perturbation
off\cite{zhao}. In Fig.3 we show the constraints on the parameters
in Eq. (2) from SNIa, WMAP and SDSS and for comparison we plot in
Fig.4 the corresponding constraints without the perturbation\cite{XZ}.

\begin{figure}[htbp]
\includegraphics[scale=0.3]{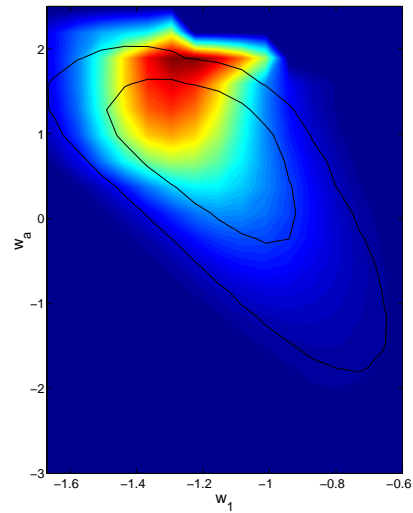}
\caption{Constraints on parameters of model (2) from SNIa + WMAP +
SDSS with dark energy perturbation\cite{XZ}.}
\end{figure}

\begin{figure}[htbp]
\includegraphics[scale=0.3]{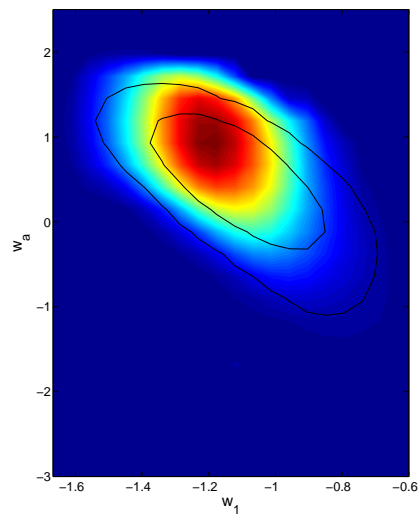}
\caption{Constraints on parameters of model (2) from SNIa + WMAP +
SDSS without dark energy perturbation\cite{XZ}.}
\end{figure}

\section{A new scenario of dark energy: the Quintom}

Currently the cosmological constant fits well to the data,
however observational data
do not exclude the possibility of dynamical dark energy. Actually
a class of models where
the equation of state or an effective EOS evolves and
crosses
over the cosmological constant boundary is mildly favored.
If such a result holds on with the accumulation
of observational data, this would be a great challenge to current
cosmology.
 Firstly, the cosmological constant as a
candidate for dark energy will be excluded and dark energy must be
dynamical. Secondly, the simple dynamical dark energy models
considered vastly in the literature like the Quintessence, the
Phantom or the K-essence\cite{Vikman} can not be satisfied either.

 A simple Quintom model consists of
two scalar fields, one being the Quintessence with the other being
the Phantom field\cite{FWZ}. This type of Quintom model will
provide a scenario where at early time the Quintessence dominates
with $w >-1$ and lately the Phantom dominates with $w$ less than
$-1$, satisfying current observations. A detailed study on the
cosmological evolution of this class of Quintom model is performed
in Ref.\cite{guo}. The Quintom models are different from the
Quintessence or Phantom in the determination of the evolution and
fate of the universe. Generically speaking, the Phantom model has
to be more fine tuned in the early epochs to serve as dark energy
today, since its energy density increases with expansion of the
universe. Meanwhile the Quintom model can also preserve the
tracking behavior of Quintessence, where less fine tuning is
needed.

In addition to the Quintom model mentioned above there are at
least two more possibilities in the Quintom model buildings.
 One will be the scalar field models with
non-minimal coupling to the gravity
where the effective equation of the state can be arranged to
change from above $-1$ to below $-1$ and vice versa. For a single
scalar field coupled with gravity minimally, one may consider a
model with a non-canonical kinetic term with the following
effective Lagrangian\cite{FWZ}:
 \begin{equation}
\mathcal{L}=\frac{1}{2}f(T)\partial_{\mu}Q\partial^{\mu}Q-V(Q)~,
 \end{equation}
where $f(T)$ in the front of the kinetic term is a dimensionless
function of the temperature or scalar fields. During the evolution
of the universe when $f(T)$ changes sign from positive to negative
it gives rise to an realization of the interchanges between the
Quintessence and the Phantom scenarios.

Since last year there have been a lot of studies in the literature
on the Quintom-like model building\footnote{I apologize for not
refereing to the papers relevant to this talk due to the
limitation on the length of the paper.}. In general as argued in
Ref.\cite{zhao} in the conventional case with an single perfect
fluid or a single scalar field one will not be able to realize a
viable Quintom model, so to have the $w$ crossing $-1$ one needs to
introduce extra degree of freedom. One possibility is to include
multi fluids or multi scalar fields as shown above. Another
possibility is to introduce the higher derivative operators to the
lagrangian. This is the model proposed in Ref. \cite{MFZ05} by
introducing higher derivative operators to the Lagrangian.
Specifically in \cite{MFZ05} we considered a model with the
Lagrangian
 \be\label{lagrangian} \mathcal{L}=-{1\over
2}\nabla_{\mu}\phi\nabla^{\mu}\phi+{c\over
2M^2}\Box\phi\Box\phi-V(\phi)~, \ee where $\Box\equiv
\nabla_{\mu}\nabla^{\mu}$ is the d'Alembertian operator. The term
related to the d'Alembertian operator is absent in the
Quintessence, Phantom and the k-essence model, which is the key to
make the model possible for $w$ to cross over $-1$. We have proven
in \cite{MFZ05}
 this
Lagrangian is equivalent to an effective two-field
model\be\label{alagrangian} \mathcal{L}= -{1\over
2}\nabla_{\mu}\psi\nabla^{\mu}\psi+{1\over
2}\nabla_{\mu}\chi\nabla^{\mu}\chi -V(\psi-\chi)-{M^2\over
2c}\chi^2~, \ee with the following definition\bea
\chi &=&\frac{c}{M^2}\Box\phi\label{change}~,\\
\psi &=&\phi+\chi~. \eea Note that the redefined fields $\psi$ and
$\chi$ have opposite signs in their kinetic terms. One might be able
to derive the higher derivative term in the effective Lagrangian
(5) from fundamental theories. In fact it has been shown in the
literature that this type of operator does appear as some quantum
corrections or due to the non-local physics in the string theory.
It is interesting and worthwhile to
study further the implications of models with higher derivatives in
cosmology (for a recent study see e.g. \cite{relevnt1}).

In the following I summarize some of the interesting aspects associated
with
the Quintom scenario of dark energy:

1) Quintom dark energy gives rise to a new scenario of the evolution and
the
fate
of the Universe. For example,
in Ref.\cite{paper10}we have studied a class of Quintom models
with an oscillating equation of state and found that
oscillating Quintom can unify the early inflation and current
acceleration of the universe, leading to oscillations of the
Hubble constant and a recurring universe. Our oscillating Quintom
would not lead to a big crunch nor big rip. The scale factor keeps
increasing from one period to another and leads naturally to a
highly flat universe. The universe in this model recurs itself and we are
only staying among one of the epochs, in which sense the
coincidence problem is reconciled.

2) The study on the Quintom models provide us a way to include the
perturbation of the dark energy consistently\cite{zhao}, which is
important when performing the global fittings of the parametrized
equation of state to the observational data including CMB,
LSS and so on. As I pointed out in section II, neglecting the perturbation
in general will lead to some bias in the fitting results.

3) The Quintom model might provide a way to solving the problem of
the quantum instability inherited in the model of
Phantom\cite{trodden}. For Phantom theory, the equation of state
is always less than $-1$, however in the scenario of Quintom, $w <
-1$ happens just for a short period of time very much like the
"tachyon" existing only during the phase transition. It might be
possible that the quantum instability problem be solved in the
context of the higher derivative theory of Quintom model shown in
(5). As pointed out in Ref.\cite{hawking}, the problem arises
because $\chi$ and $\psi$ in Eq.(6) are quantized in canonical way
independently. In fact, both of them are determined by $\phi$. A
more appropriate quantization method seems to be possible to avoid
the instability.

\section{Neutrino dark energy}

Recently there have been a lot of interests in the
literature[18-35] in studying the possible connections between the
neutrinos and the dark energy.
 There are at least two observations which
motivate these studies: 1) the dark energy scale $\sim 10^{-3}$ eV
is smaller than the energy scales in particle physics, but
interestingly is comparable to the neutrino masses; 2) in
Quintessence-like models of dark energy $m_Q \sim 10^{-33}$ eV,
which surprisingly is also connected to the neutrino masses {\it
via} a see-saw formula $m_Q \sim {m_\nu^2 / M_{Pl}}$ with $M_{Pl}$
the planck mass.

Are there really any connections between the neutrinos and dark
energy? Given the arguments above it is quite interesting to make
such a speculation on this connection. If yes, however in terms of
the language of the particle physics it requires the existence of
new dynamics and new interactions between the neutrinos and the
dark energy sector.

Qualitatively these models have made at least two interesting
predictions: 1) neutrino masses are not constant, but vary during
the evolution of the universe; 2) CPT is violated in the neutrino
sector due to the CPT violating {\bf Ether} during the evolution
of the Quintessence scalar field\cite{paper11, paper15}.
Quantitatively these predictions will depend on the dynamics
governing the coupled system of neutrino and dark energy.
  Here I focus on the models with mass varying neutrinos.

In general for the models of neutrino dark energy or interacting
dark energy, the lagrangian can be written as
\begin{eqnarray}
{\cal L}= {\cal L}^{SM}_{\nu} +  {\cal L}_{Q} + {\cal L}_{int},
\end{eqnarray}
where ${\cal L}^{SM}_{\nu}$ is the lagrangian of the standard
model (SM) describing the physics of the left-handed neutrinos,
${\cal L}_{Q}$ is for the dynamical scalar such as Quintessence.
${\cal L}_{int}$ in (9) is the sector which mediates the
interaction between the dark energy scalar and the  neutrinos.

At energy much below the electroweak scale, the relevant lagrangian for
the neutrino dark energy is given by
\begin{equation}
{\cal L}={\cal L}_\nu + {\cal L}_Q +M(Q)\bar{\nu}{\nu}\ ,
\end{equation}
where ${\cal L}_\nu$ is the kinetic term of the neutrinos.
The last term of Eq.(10) is the scalar field dependent mass of the
neutrinos which characterizes the interaction between the
neutrinos and the dark energy scalar. In the standard model of
particle physics, the neutrino masses can be described by a
dimension-5 operator
  \begin{equation}
  L_{\not L}=\frac{2}{f}l_{L}l_{L}HH+h.c,
  \end{equation}
  where $f$ is a scale of new physics beyond the Standard Model
  which generates the $B-L$ violations, $l_{L},  H$ are the
 left-handed lepton and Higgs doublets respectively. When the
  Higgs field gets a vacuum expectation value $<H> \sim v$, the
  left-handed neutrino receives a majorana mass
  $m_{\nu} \sim \frac{v^{2}}{f}$.
 In Ref.\cite{paper31} we considered
an interaction between the neutrinos and the Quintessence $Q$
 \begin{equation}
 \beta \frac { Q }{M_{Pl}} \frac{2 }{f} l_{L}l_{L}
HH+ h.c  ,
 \end{equation}
 where $\beta $ is the coefficient which characterizes the strength of the
Quintessence interacting with the neutrinos. In this scenario the
neutrino masses vary during the evolution of the universe and we
have shown that the neutrino mass limits imposed by the
baryogenesis are modified.

The dim-5 operator above is not renormalizable, which in principle
can be generated by integrating out the heavy particles. For
example, in the model of the minimal see-saw mechanism for the
neutrino masses,
\begin{equation}
L=h_{ij}\bar{l}_{Li}N_{Rj}H+\frac{1}{2}M_{ij}\bar{N}^{c}_{Ri}N_{Rj}+h.c.
 \end{equation}
 where $ M_{ij}$ is the mass matrix of
the right-handed neutrinos and the Dirac masses of the neutrinos are given
by $m_{D}\equiv h_{ij} <H> $. Integrating out the heavy
right-handed neutrinos one will generate a dim-5 operator, however as
pointed out in Ref.\cite{paper31} to have the light neutrino
masses varied there are various possibilities, such as by coupling
the Quintessence field to either the Dirac masses or the majorana
masses of the right-handed neutrinos or both.

In Ref.\cite{paper33} we have specifically proposed a model of
mass varying right-handed neutrinos. In this model the
right-handed neutrino masses $M_i$ are assumed to be a function of
the Quintessence scalar $M_i(Q)=\overline{M}_i e^{\beta
\frac{Q}{M_{Pl}}}$. Integrating out the right-handed neutrino will
generate a dimension-5 operator, but for this case the light
neutrino masses will vary in the following way
\begin{equation}
 e^{-\beta
\frac{Q}{M_{Pl}}} \frac{2 }{f} l_{L}l_{L} HH+ h.c  .
 \end{equation}

Some interesting features of the neutrino dark energy models can
be summarized as follows:

1) With mass varying the neutrinos become a part of the dark energy
and play an different role in the determination of the evolution of the
Universe than the
traditional
non-relativistic matter.
 In Ref. \cite{BFLZ} we have studied in detail
the cosmological evolution of the
Universe in the scenario of mass varying neutrinos. We found that
the neutrino density will not decrease and interestingly
become the dominant one for a suitable choice of the model parameters.
Mass varying neutrinos have interesting implications in
leptogenesis\cite{paper31, paper33}.

2) The predictions on the variation of the neutrino masses can be tested
with Short Gamma
Ray Burst\cite{grb}, CMB and LSS\cite{nucmb} and much more interestingly
and importantly in the experiments of neutrino oscillation\cite{paper35,
paper37}.

3) Neutrinos coupled to dark energy scalar\cite{twofield} can provide a
scenario of dark energy with
the equation of state crossing the cosmological constant boundary of
$-1$. In a recent paper\cite{hongli} we used the recently released  SNIa
data to
constrain the couplings between the neutrinos and the dark energy
scalar. We found the current data mildly favor the model
where the Phantom-like scalar couples to the neutrinos.

\section{Discussions and conclusion}

In this talk I have firstly given a brief review on the
constraints on dark energy from observational data and presented a
review on the Quintom model of dark energy. Secondly I have
discussed the interacting dark energy and focused on the possible
interactions between the dark energy and the neutrinos. If these
interactions indeed exist, they will open up some possibilities of
detecting the dark energy non-gravitationally. Before concluding I
wish to add a new possibility to the list of interacting dark
energy models\cite{xmzhang} where the dark energy sector is
closely connected to the Higgs and the Top quark in the standard
model of elementary particle physics(SM). The motivation for this
study is the following: the dark energy scale $\Lambda_{DE} \sim
10^{-3}$ eV is numerically comparable to $\Lambda_{DE} \sim
\frac{\Lambda_{F}^2}{M_{Pl}}$ where $\Lambda_{F}$ is the Fermi
scale, $M_{Pl}$ the Planck mass. In the particle physics the
sector of the electroweak symmetry breaking includes the Higgs
boson. The top quark since its mass is heavier than all of the
particles observed so far is believed to strongly couple to and
play an essential role in understanding the physics associated to
the Fermi scale. For example, in the models of top quark
condensation, the Higgs is composite of the top pairs. Motivated
by the observations above and numerically the masses of the top
quark and the Higgs boson are order of the Fermi scale, the
Quintessence scalar might preferably couple to the top quark and
the Higgs than to other light particles.
 In this model
the Higgs mass
and the CP violating phase associated with the top Yukawa
coupling vary during the evolution of the Quintessence. I will show
that in this model
the first order phase transition will be strong enough for
electroweak baryogenesis without
conflicting the current experimental limit on the Higgs mass.

In the SM, the Higgs potential at the tree level is given by
\begin{eqnarray}
V(H) = \lambda {[H^+ H - \frac{v^2}{2}]}^2,
\end{eqnarray}
where the $\lambda$ is the self coupling constant of the Higgs
field. When the $SU(2)$ doublet Higgs fields $H$ gets a vacuum
expectation value $v$, the physical Higgs boson $h$ receives a
mass $m_h = 2 \lambda v^2$. Now we assume the interaction between
the Higgs field $H$ and the Quintessence scalar $Q$ explicitly to
be
\begin{eqnarray}
\beta \lambda \frac{(Q-Q_0)}{M_{Pl}}{[H^+ H - \frac{v^2}{2}]}^2,
\end{eqnarray}
where $Q_0$ is the value of the Quintessence field at present time
and the parameter $\beta$ characterizes the strength of this type
of interaction. Combining (15) and (16) we obtain an effective self
coupling of the Higgs field
\begin{eqnarray}
{\lambda}^{eff} = \lambda [ 1 + \beta \frac{(Q-Q_0)}{M_{Pl}}].
\end{eqnarray}
From (17) one can see the Higgs mass $m_h^2 = 2 \lambda^{eff} v^2$
is now a function of the Quintessence field. At present time
$m_h^2 = 2 \lambda v^2$ which recovers the results in the
literature. At early time of the Universe, however $Q$ differs
from $Q_0$, consequently the Higgs mass will also differ from its
present value. This opens a possibility of having a light Higgs
during the electroweak phase transition without conflicting the
experimental limit at present time\footnote{this possibility has
been mentioned in Ref\cite{paper31}.}.
 Quantitatively the amount of the changes in the Higgs mass depends on
the evolution of the Quintessence field. For a specific discussion
we consider a model of Quintessence with a inverse power-law
potential $V(Q) = V_0Q^{-\alpha}$\cite{model1}. In Fig. 5 we plot
the evolution of the Quintessence scalar as a function of $\ln a$.
One can see that in the early time during the period of  the
radiation dominate or matter dominate the scalar field was almost
frozen and the value of $Q$ is nearly a constant, then change to
the present value very recently. In Fig. 6 we plot the Higgs mass
as a function of the $\ln a$ from which one can see that with a
Higgs mass well above the current experimental limit $m_h > 115$
GeV, at temperature around 100 GeV it could be as light as $35$
GeV with $\beta \sim 5$.

\begin{figure}[htbp]
\includegraphics[scale=0.6]{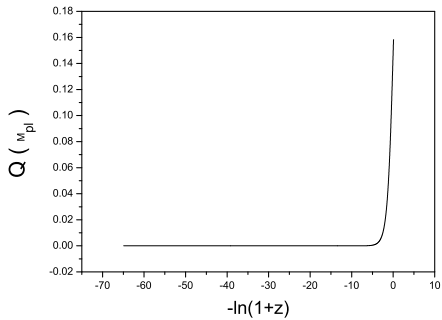}
\caption{Plot of the evolution of the scalar field of Quintessence
with the inverse power-low potential $V(Q) = V_0Q^{-\alpha}$ where
 $\alpha=0.5$.}
\end{figure}
\begin{figure}[htbp]
\includegraphics[scale=0.6]{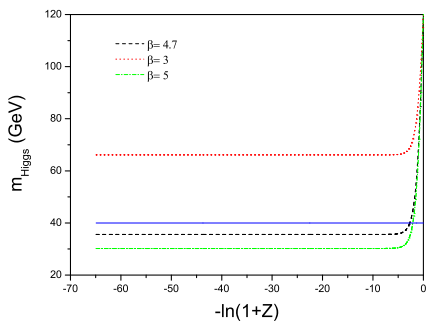}
\caption{Plot of Higgs mass with different value of parameter
$\beta$ where $\beta=4.7$ (dashed line), $\beta=3$ (dotted line)
and $\beta=5$ (dash dotted line); the horizontal line indicates
 the mass of the Higgs at 40 GeV.}
\end{figure}

Introducing the coupling between the Quintessence scalar and the top quark
will cause the top Yukawa coupling running.
In the SM the top quark Yukawa coupling is
\begin{eqnarray}
y_t \bar{\Psi_L}{\tilde H} t_R,
\end{eqnarray}
where $\Psi_L = (t_L, b_L)$ is the SU(2) doublet quark of the
third generation. We introduce a coupling of the Quintessence to
the term in (18)
\begin{eqnarray}
 \delta \frac{(Q-Q_0)}{M_{Pl}}\bar{\Psi}_L{\tilde H} t_R,
\end{eqnarray}
where $\delta$ is a parameter which characterizes this effective
interaction. In general the Yukawa coupling in (18) and the
coupling $\delta$ in (19) are complex with different phases. By the
redefinition of the quark field and the convenience of the
discussion we assume $y_t$ is real and $\delta$ has a phase.
Introducing explicitly a phase $\xi$ and defining $\delta = c_t y_t
e^{i \xi}$ we have an effective Yukawa coupling of the Top quark
by combining (18) and (19)
\begin{eqnarray}
y_t^{eff} = y_t [ 1 + c_t e^{i \xi} \frac{(Q-Q_0)}{M_{Pl}} ].
\end{eqnarray}
One can see from the equation above that at present time the top
quark mass $m_t = y_t v / \sqrt{2}$ is real, however at early time
of the Universe it includes a complex CP violating phase.
Unfortunately since the Quintessence field is almost a constant
during the electroweak phase transition, the CP violation in the
effective top Yukawa coupling will not help for the generation of
the baryon number asymmetry\footnote{I thank Tim Tait and L.
Fromme for discussions.}.

\section{acknowledgements}
I would like to thank the organizers of PASCOS 2005 for invitation and
giving me the opportunity to present this talk. I thank my collaborators
for discussions. This work is
supported in part by National Natural Science Foundation of China
(grant Nos. 90303004, 19925523) and by Ministry of Science and
Technology of China( under Grant No. NKBRSF G19990754).

\end{document}